\documentclass{llncs}

\usepackage{makeidx}  

\usepackage{amsmath}

\usepackage{mathptmx}

\usepackage{SIunits}            
\usepackage{courier}            
\usepackage[scaled]{helvet} 
\usepackage{url}                  
\usepackage{listings}          
\usepackage{enumitem}      
\usepackage{times}
\usepackage{xspace}
\usepackage{url}

\usepackage{todonotes}
\usepackage{alltt}
\usepackage{balance}
\usepackage{graphicx}

\newcommand{\chess}{\textsc{Chess}\xspace}
\newcommand{\vcc}{\textsc{Vcc}\xspace}
\newcommand{\XSpace}[1]{}

\newcommand{\squishlist}{
   \begin{list}{$\bullet$}
    { \setlength{\itemsep}{1.5pt}      \setlength{\parsep}{3pt}
      \setlength{\topsep}{3pt}       \setlength{\partopsep}{1.5pt}
      \setlength{\leftmargin}{1.5em} \setlength{\labelwidth}{1em}
      \setlength{\labelsep}{0.5em} } }

\newcommand{\squishend}{
    \end{list}  }

 \newcommand{\squishdes}{
   \begin{description}{$\bullet$}
    { \setlength{\itemsep}{0pt}      \setlength{\parsep}{3pt}
      \setlength{\topsep}{3pt}       \setlength{\partopsep}{0pt}
      \setlength{\leftmargin}{1.5em} \setlength{\labelwidth}{1em}
      \setlength{\labelsep}{0.5em} } }

\newcommand{\squishdesend}{
    \end{description}  }

\lstdefinelanguage{CRDT} {
    morekeywords={content,initial,query,pre,let,update,atCoordinator,downstream},
    moredelim=[is][\itshape]{|}{|}
}

\begin{document}




\title{Inferring Formal Properties of\\ Production Key-Value Stores}
\titlerunning{Inferring Formal Properties of Production Key-Value Stores}

\author{Edgar Pek \and Pranav Garg \and Muntasir Raihan Rahman \and\\ Karl Palmskog \and Indranil Gupta \and P.~Madhusudan}
\tocauthor{Pek et al.}
\institute{University of Illinois at Urbana-Champaign, Urbana IL 61801, USA\footnote{This work was done by the authors when they were at Illinois. Pek is now at Adobe, Garg is now at Amazon, and Rahman is now at Microsoft.}\\
  \email{\{pek1,garg11,mrahman2,palmskog,indy,madhu\}@illinois.edu}}

\maketitle
\begin{abstract}
  Production distributed systems are challenging to formally verify, in particular when they are
  based on distributed protocols that are not rigorously described or fully understood. In this paper,
  we derive models and properties for two core distributed protocols used in eventually
  consistent production key-value stores such as Riak and Cassandra.
  We propose a novel modeling called certified program models,
  where complete distributed systems are captured as programs written in traditional systems
  languages such as concurrent C.
  Specifically, we model the read-repair and hinted-handoff recovery protocols as concurrent C programs, test them
  for conformance with real systems, and then verify that they guarantee eventual consistency,
  modeling precisely the specification as well as the failure assumptions under which the results hold.
\end{abstract}

\section{Introduction}

Distributed systems are complex software systems that pose myriad challenges to formal verification. Some systems are constructed from rigorously described distributed algorithms~\cite{Paxos}, which requires bridging a substantial gap from an abstract algorithm to executable code~\cite{PaxosMadeLive}. The implementation of distributed protocols developed this way is, of course, not usually formally proven
to be a refinement of the textbook algorithm, though some research on developing the implementation using \emph{formally verified refinements} has been done~\cite{Hawblitzel15}. However, most \emph{production systems} have not been built through a top-down approach from well-understood and proven-on-paper algorithms, but rather have been developed in an ad-hoc fashion, from scratch (e.g., on whiteboards), undergoing iteration and revision over a long period of time. A large bulk of today's open-source distributed systems software fits this category, the most prominent among these being key-value/NoSQL storage systems. 
Consequently, to understand the formal properties guaranteed by these systems, we need to build
high-level protocol models and infer the properties guaranteed by these protocols using the implementation and its available documentation.

In this paper, we build models and using them derive formal properties for two core distributed protocols used in \emph{eventually consistent} distributed key-value stores: the \emph{hinted-handoff protocol} and the \emph{read-repair} protocol. Eventually consistent key-value stores originate from the Dynamo system by Amazon~\cite{DeCandia07} and are currently implemented in production key-value stores such as Cassandra\footnote{\url{http://cassandra.apache.org}} and Riak\footnote{\url{http://basho.com/products/riak-kv/}}. They are used widely today, e.g., Amazon relies on Dynamo for its shopping cart, and Cassandra is used by Netflix and many other companies. Yet, none of these systems were built from rigorously proven distributed algorithms.

Our approach is to model the high-level protocols in these implementations, and then derive formal properties that the models guarantee
by finding properties that are formally verifiable for the models. We derive formal properties for the two protocols mentioned above and
verify them (against the model) by using a novel methodology called \emph{certified programs models}, where a high-level distributed algorithm is modeled using \emph{programs} written in traditional systems languages, e.g., C with concurrency,
and then certified to be correct against their specifications using program verification. The program models capture not only the behavior of distributed processes and their memory and secondary storage states, but also network communication, delays, and failures, using non-determinism where necessary.

Modeling and verification using certified program models has several salient aspects.
First, program models are \emph{executable} and can be validated for conformance to the system using testing,
where the programmer can write test harnesses that control inputs as well as physical events such as node and network failures,
and test using mature systematic testing tools for concurrent software, e.g., \chess~\cite{Musuvathi08}.
Moreover, program models permit accurate modeling of specifications of protocols using \emph{ghost state} as well as
assertions in powerful logics. Finally, program models lend themselves well to \emph{program verification techniques}, especially using tools such as \vcc~\cite{Cohen09} that automate large parts of the reasoning using logical constraint solvers.
Our experience in this work shows that certified programs models are an appealing sweet spot for verifying distributed prototcols,
that facilitates executable models that capture arbitrarily parameterized protocols and at the same time are amenable to mostly automated verification.
\vspace{-0.5cm}
\subsection{Key-value Stores and Eventual Consistency}

Key-value storage systems arose out of the CAP theorem/conjecture, which was postulated by Brewer and proved by Gilbert and Lynch~\cite{lg:cap}. The conjecture states that a distributed storage system can choose at most two out of three important char\-ac\-te\-ris\-tics---strong data Consistency (i.e., linearizability or sequential consistency), Availability of data (to reads and writes), and Partition-tolerance. Hence, achieving strong consistency while at the same time providing availability in a partitioned system with failures is impossible.
While traditional databases preferred consistency and availability, the new generation of key-value systems are designed to be partition-tolerant, both within a datacenter as well as across multiple data-centers. As a result, a key-value system is forced to chose between one of either strong consistency or availability---the latter option provides lower latencies for reads and writes~\cite{Abadi12}.
Key-value systems that prefer availability include Cassandra, Riak, and Dynamo~\cite{DeCandia07}, and support weak models of consistency (e.g., eventual consistency). Other key-value systems, e.g., Bigtable~\cite{DBLP:journals/tocs/ChangDGHWBCFG08}, instead prefer strong consistency, and may be unavailable under failure scenarios.

One popular weak consistency notion is eventual consistency, which roughly speaking, says that {\it if no further updates are made to a given data item, all replicas will eventually hold the same value (and a read would then produce this value)}~\cite{DeCandia07}. Eventual consistency is a \emph{liveness property}, not a safety property~\cite{Bailis:2013:ECT:2460276.2462076}.
The precise notion of what eventual consistency means in these protocols (the precise assumptions under which they hold, the failure models, the
assumptions on the environment, etc.) are not well understood, let alone proven. Programmers also do not understand the subtleties of eventually
consistent stores; for instance, default modes in Riak and Cassandra can permanently lose writes---this has been exploited
in an attack involving Bitcoin\footnote{\url{http://hackingdistributed.com/2014/04/06/another-one-bites-the-dust-flexcoin/}}.
\vspace{-0.3cm}
\subsection{Contributions}
\label{sec:contribs}

The primary contribution of this paper is to precisely reason about the guarantees of eventual consistency that
two core protocols used in production implementations of key-value stores provide.
More specifically, we model and verify the correctness of the \emph{hinted-handoff} protocol and the \emph{read-repair} protocol, which are anti-entropy mechanisms first proposed in the Amazon Dynamo system~\cite{DeCandia07}, and later implemented in systems such as Riak and Cassandra.

We build program models for these protocols in concurrent C that we verify for eventual consistency.
The programs use \emph{threads} to model concurrency, where each get/put operation as well as the asynchronous calls they make
are modeled using concurrently running threads.
The state of the processes, such as stores at replicas and the hinted-handoff tables, are modeled
as shared arrays. Communication between processes is also modeled using data-structures: the network is simulated using
a set that stores pending messages to replicas, with an independent thread sending them to their destinations.
Failures and non-determinism of message arrivals, etc., are also captured programmatically using
non-determinism (modeled using stubs during verification and using random coin-tosses during testing).
In particular, system latency is captured by background threads that are free to execute any time,
modeling arbitrarily long delays.

In the case of the \emph{hinted-handoff protocol}, we prove that this protocol working alone guarantees eventual consistency provided there
are only transient faults. In fact, we prove a stronger theorem by showing that for any operation based (commutative) conflict-free replicated data-type implementing a register, the protocol ensures \emph{strong eventual consistency}--- this covers a variety of schemes that systems use, including Riak and Cassandra, to resolve conflict when implementing a key-value store. Strong eventual consistency guarantees not only eventual consistency, but that the store always contains a value which is a function of the set of updates it has received, independent of the order in which it was received. We prove this by showing that the hinted-handoff protocol (under only transient failures) ensures
\emph{eventual delivery} of updates---when this is combined with an idempotent and commutative datastructure like a CmRDT~\cite{Shapiro-Tech-Report11} , it ensures strong eventual consistency.
We model the eventual delivery property in the program model using a ghost \emph{taint} that taints a particular write at a coordinator
(unbeknownst to protocol), and asserts that the taint propagates eventually to every replica.
Like eventual consistency, eventual delivery is also a \emph{liveness property}. It is established by finding a ranking function
that models abstractly the time needed to reach a consistent state, and a slew of corresponding safety properties to prove this program correct.

For the \emph{read-repair protocol}, the literature and documentation of the above systems indicated
that a read-repair (issued during a read) would bring the nodes
that are alive to a consistent state eventually. However, while attempting to prove this property, we realized that
no invariant could prove this property, and that it is false.
In fact, a single read is insufficient to reach eventual consistency. Hence, we prove a more complex property: beyond a point, if a set of nodes are alive and they all
stay alive, and if all requests stop except for an unbounded sequence of reads to a key, then the live nodes that are responsible
for the key will eventually converge. In other words, one read is not enough for convergence, and the system needs a long series of reads.

Note that the certification that the program models satisfy their specification
is for an \emph{unbounded} number of threads, which model an unbounded number of replicas,
keys, values, etc., model arbitrarily long input sequences of updates and reads to the keys, and model the concurrency prevalent in the system
using parallelism in the program. The verification is hence a \emph{complete} verification of the models, in contrast to approaches using under-approximations to systematically test a bounded-resource system~\cite{maudeCass,Newcombe15}. \XSpace{In particular, Amazon has reported modeling of distributed protocols using TLA, a formal system, and used model-checking (systematic testing) on bounded instances of the TLA system to help understand the protocols, check their properties, and help make design decisions.}  Our approach is to model protocols using C programs, which we believe are much simpler for systems engineers to use to model protocols, and being executable, are easy to test using test harnesses. Most importantly, we have proved the entire behavior of the protocol correct \XSpace{(as opposed to the work using TLA)}using the \XSpace{state-of-the-art} program verification framework \vcc~\cite{Cohen09} that automates several stages of the reasoning.

\paragraph{Paper Organization.} The rest of the paper is structured as follows. Section~\ref{sec:bg} gives more details on key-value stores, eventual consistency, and the read-repair and hinted-handoff anti-entropy protocols\XSpace{ (readers familiar with these topics can choose to skip this section)}. We state our main results in Section~\ref{sec:main-result}, where we describe the precise properties we prove for the protocol models as well as some properties that we expected to be true initially, but which we learned were not true in general. Section~\ref{sec:model} describes our program models of protocols in detail, including the testing approach we used to check that our model was reasonable. The verification process, including background on program verification, the invariants and ranking functions required for proving the properties is covered in Section~\ref{sec:verification}. Section~\ref{sec:related} describes related work and Section~\ref{sec:conclusion} concludes.

\section{Background}
\label{sec:bg}
In this section, we describe the read and write paths involved in a key-value store, and the anti-entropy protocols which are used to implement eventual consistency by reconciling divergent distributed replicas.\XSpace{ Readers familiar with key-value store system internals can skip this section without loss of continuity.} Key-value stores persist pairs of keys and values, and usually have two basic operations: get(key) for retrieving the value corresponding to a key, and put(key, value) for storing the value of a particular key\XSpace{\footnote{We use both read/write and get/put terms to mean data fetch and data update operations.}}. Key-value stores typically use consistent hashing~\cite{Karger:1997:CHR:258533.258660} to distribute keys to servers, and each key is replicated across multiple servers for fault tolerance. When a client issues a put or get operation, it first interacts with a server, e.g., the server closest to the client. This server acts as a \emph{coordinator}: it coordinates the client and replica servers to complete the put and get operations. The CAP theorem~\cite{lg:cap} implies that during network partitions (where servers are split into two groups with no intercommunication), a key-value store must choose either strong consistency (linearizability)~\cite{her:lin} or availability. Even when the network is not partitioned, the system is sometimes configured to favor latency over consistency~\cite{abadi:cap}. As a result, popular key-value stores \XSpace{like Cassandra~\cite{laksh:cass} and Riak~\cite{riak}}expose tunable \emph{consistency levels}.
These consistency levels control the number of servers the coordinator needs to hear from before declaring success on reads and writes.
For instance, a write threshold of one allows the system to return with success after writing to just one replica.
When the sum of read and write thresholds is greater than the number of replicas, the system will ensure strong consistency.

In general, a consistency model can be characterized by its restrictions on operation ordering. The strongest models, e.g., linearizability~\cite{her:lin} severely restrict possible operation orderings that can lead to correct behavior. Eventual consistency, in contrast, is one of the weakest consistency models. Informally, it guarantees that, if no further updates are made to a given data item, reads to that item will eventually return the same value~\cite{V09}. Thus, until some undefined time in the future when the system is supposed to converge, the user can never rule out the possibility of data inconsistency.\XSpace{Despite the weak guarantees, many applications have been successfully built on top of eventually consistent stores.}

To achieve high availability and reliability, key value stores typically replicate data on multiple servers. For example, each key can be replicated on $N$ servers, where $N$ is a configurable parameter. In the weakest consistency setting (with read and write thresholds of one), each get and put operation only touches one replica (e.g., the one closest to the coordinator). Thus, in the worst case scenario where all puts go to one server, and all get operations are served by a different server, the replicas will never converge to the same value. To ensure convergence to the same value, production key-value stores employ \emph{anti-entropy} protocols~\cite{Demers:1987:EAR:41840.41841}. An anti-entropy protocol operates by comparing replicas and reconciling differences. The three main anti-entropy protocols are read-repair, hinted-handoff, and node-repair. While the first two are \emph{real-time} protocols involved in, respectively, the read and write paths, the third one is an offline protocol, which runs periodically\XSpace{ (e.g., during non-peak load hours)} to repair out-of-sync nodes (e.g., when a node rejoins after recovering from a crash). Here, we only consider the real-time anti-entropy protocols. \XSpace{Node-repair is mostly an offline process whose correctness lies solely in the semantics of the merge, so we do not consider it in this paper.}

Read-repair~\cite{DeCandia07} is a real-time anti-entropy mechanism that ensures that all replicas have (eventually) the most recent version of a value for a given key.
In a typical read path, the coordinator forwards read requests to all replicas, and waits for a consistency level ($CL$ out of $N$) number of replicas to reply. If read-repair is enabled, the coordinator checks all the read responses (from the nodes currently alive), determines the most recent read value\XSpace{\footnote{Determining the most recent version of data to push to out of date replicas is implementation dependent. For Cassandra, the replica value with highest client timestamp wins. Riak uses vector clocks to decide the winner, and can deduce multiple winners in case of concurrent writes.}}, and finally pushes the latest version to all out of date replicas.

Hinted-handoff~\cite{DeCandia07}, unlike read-repair, is part of the write path. It offers full write availability in case of failures, and can improve consistency after temporary network failures. When the coordinator finds that one of the replicas responsible for storing an update is temporarily down (e.g., based on failure detector predictions), it stores a hint meta-data for the down node for a configurable duration of time. Once the coordinator detects that the down node is up, it will attempt to send the stored hint to that recovered node. Thus hinted-handoff ensures that no writes are lost, even in the presence of temporary node failures. In other words, this mechanism is used to ensure that eventually all writes are propagated to all the replicas responsible for the key.

\section{Characterizing and Proving Eventual Consistency}
\label{sec:main-result}

\XSpace{The goal of this paper is to prove eventual consistency of the hinted-handoff
and read-repair protocols that systems like Cassandra and Riak implement,
delineating precisely the conditions under which they hold. Our effort spanned a period of 15 months, with about 6 person months of effort for modeling and verification.}

To model and verify the read-repair and hinted-handoff protocols, we first \emph{abstract} away from the particular
instantiation of these protocols in these systems, and also abstract away
from the various options they provide to users to modify the behavior of the system. For instance, in Riak, using one set of options, every write is tagged with a vector clock at the client, and every replica responsible for it maps it to a \emph{set of values}, one for each last concurrent write that it has received. When a read is issued, Riak can return the set of \emph{all} last concurrently written values to it (these values are called ``siblings'' in Riak). However, in Cassandra, vector clocks are not used; instead each client labels every write with a timestamp, and despite there being drift among the clocks of clients, each replica stores only the last write according to this timestamp.\XSpace{ Furthermore, these policies can be changed; e.g., in Riak, a user can set options to mimic the Cassandra model.}

We capture these instantiations by generalizing the semantics of how the store is maintained. For the hinted-handoff protocol, we prove eventual consistency under the assumption that the stores are maintained using some \emph{idempotent} operation-based commutative replicated data-type (CRDT)~\cite{Shapiro-Tech-Report11,Shapiro11} that implements a \emph{register}, while for read-repair, we prove eventual consistency assuming an arbitrary form of conflict resolution. We consider two failure modes: (a) \emph{transient failure} where failed nodes and network edges remember their pre-crash state when they come back; and (b) \emph{permanent failure} where failed nodes or network edges lose memory and start with some default state when they come back.
\XSpace{Let us first discuss the failure models we consider, which are part of the assumptions needed to prove properties of protocols.}

\subsection{Properties of the Hinted-Handoff Protocol}
\label{sec:hh}
The hinted-handoff protocol is an opportunistic anti-entropy mechanism that happens during writes. When a write is issued, and the asynchronous call to write to certain replicas fail (either explicitly or due to a timeout), the coordinator knows that these replicas could be out of sync, and hence stores these update messages in a hinted-handoff table locally to send them later to the replicas when they come back alive. However, if there is a memory wipe or a permanent failure, the hinted-handoff table will be lost, and all replicas may not receive the messages. In practice, the read-repair and node-repair protocols protect against such failures.

Our main abstraction of the key-value store is to view the underlying protocol as implementing a \emph{register} using
an operation-based conflict-free replicated datatype (CRDT)~\cite{Shapiro11}, also called a commutative replicated data-type (CmRDT). As in the work of Shapiro et al.~\cite{Shapiro-Tech-Report11}, a register is a memory cell storing opaque content. A register can be queried using a read operation get, and updated with a value $v$ using a write operation put($v$). The semantics of non-concurrent put operations corresponds to the expected sequential semantics. However, when concurrent put operations do not commute, the two common conflict resolution approaches are that (a) one operation takes precedence over the other and (b) both operations are retained. When the former approach is used, the register said to be a \emph{last write wins} (LWW) register, and when the latter approach is used, it is said to be a \emph{multi-valued} (MV) register. When implementing a simple key-value store, the vector-clock based updates in Riak can be seen as an MV register, while the simpler timestamp based updates in Cassandra can be seen as an LWW register~\cite{Shapiro-Tech-Report11}.
\XSpace{(However, since a global wall-clock time is not available, in general, this strategy in Cassandra can \emph{lose} updates~\cite{maybe-cassandra}). The CmRDTs for both LWW and MV registers are in fact idempotent---the systems tags each write with a timestamp,
and the conflict-resolution will ignore the future deliveries of
a message with same time-stamp (see~\cite{Shapiro-Tech-Report11}, Section~\ref{sec:readrepair}).}
We also assume another property of these CmRDTs, namely idempotency---we assume that all messages are tagged with a unique id, and
when a message is delivered multiple times, the effect on the store is the same as when exactly one message is delivered. Let us call such registers idempotent CRDTs\XSpace{\footnote{Standard definitions of operation-based CRDTs do not guarantee
idempotency---instead they assume the environment delivers every message
precisely once to each replica~\cite{Shapiro11}\XSpace{(see~\cite{Shapiro11}, text after Definition~5)}. Note that state-based CRDTs are usually defined to be idempotent.}}.

The main property we prove about the hinted-handoff protocol is called \emph{eventual delivery}.
This property says that every successful write eventually gets delivered to every replica at least once
(under assumptions on the kinds of possible failures and on replicas being eventually alive, etc.).
Hence, instead of eventual consistency, we argue eventual delivery, which in fact
is the precise function of these protocols, as they are agnostic of the conflict
resolution mechanism that is actually implemented in the system.
Furthermore, assuming that each replica actually implements an idempotent operation-based CRDT register,
and update procedures for these datatypes are terminating,
eventual delivery ensures eventual consistency, and in fact \emph{strong eventual consistency}~\cite{Shapiro11}.
\XSpace{(the proof in~\cite{Shapiro11} proves that when there
are reliable broadcast channels that ensure messages are delivered exactly once, CRDTs give
strong eventual consistency; this proof is readily adapted to show that when messages
are delivered at least once, idempotent CmRDTs give strong eventual consistency)}
Strong eventual consistency guarantees not only eventual consistency, but also that the store always
contains a value that is a function of the set of updates it has received, independent of the order in which they were received.

\XSpace{Our first result is that a system running only hinted-handoff-based repair provides eventual delivery of updates to all replicas, provided there are only transient faults.}
\begin{theorem}
The hinted-handoff protocol ensures eventual delivery of updates to all replicas, provided there are only transient faults. More precisely, if there is any successful write, then assuming that all replicas recover at some point, and reads and write requests stop coming at some point, the write will eventually get propagated to every replica.
\label{result1}
\end{theorem}

We formally prove the above result (and Theorem~\ref{result2} below) for arbitrary system configurations using program verification techniques on the program model; see Section~\ref{sec:model} and Section~\ref{sec:verification} for details. The following is an immediate corollary from the properties of eventual delivery and idempotent CRDTs:
\begin{corollary}
A system following the hinted-handoff protocol, where
each replica runs an operation-based idempotent CRDT mechanism that has terminating updates, is strong\-ly eventually consistent, provided there are only transient faults.\XSpace{\footnote{
The corollary may lead us to think that we can use any
operation-based CmRDT for counters at stores to obtain strong
eventually consistent counters in the presence of transient failures.
However, CmRDTs for counters are in fact \emph{not idempotent} (and
the CmRDT counters in~\cite{Shapiro11} assume that the system will
deliver messages precisely once, which hinted handoff cannot guarantee).}}
\label{corollary1}
\end{corollary}

\subsection{Properties of the Read-repair Protocol}
\label{sec:readrepair}

Our second result concerns the read-repair protocol. Read-repair is expected to be resilient to memory-crash failures, but only guarantees eventual consistency on a key provided future reads are issued at all to the key. Again, we abstract away from the conflict resolution mechanism,
and we assume that the coordinator, when doing a read and getting different replies from replicas, propagates
\emph{some} consistent value back to all the replicas.
This also allows our result to accommodate anti-entropy mechanisms~\cite{Demers:1987:EAR:41840.41841} that are used instead of read-repair, in a reactive manner after a read. Note that this result holds irrespective of the hinted-handoff protocol being enabled or disabled.

\begin{figure}
\begin{center}
\includegraphics[scale=0.2]{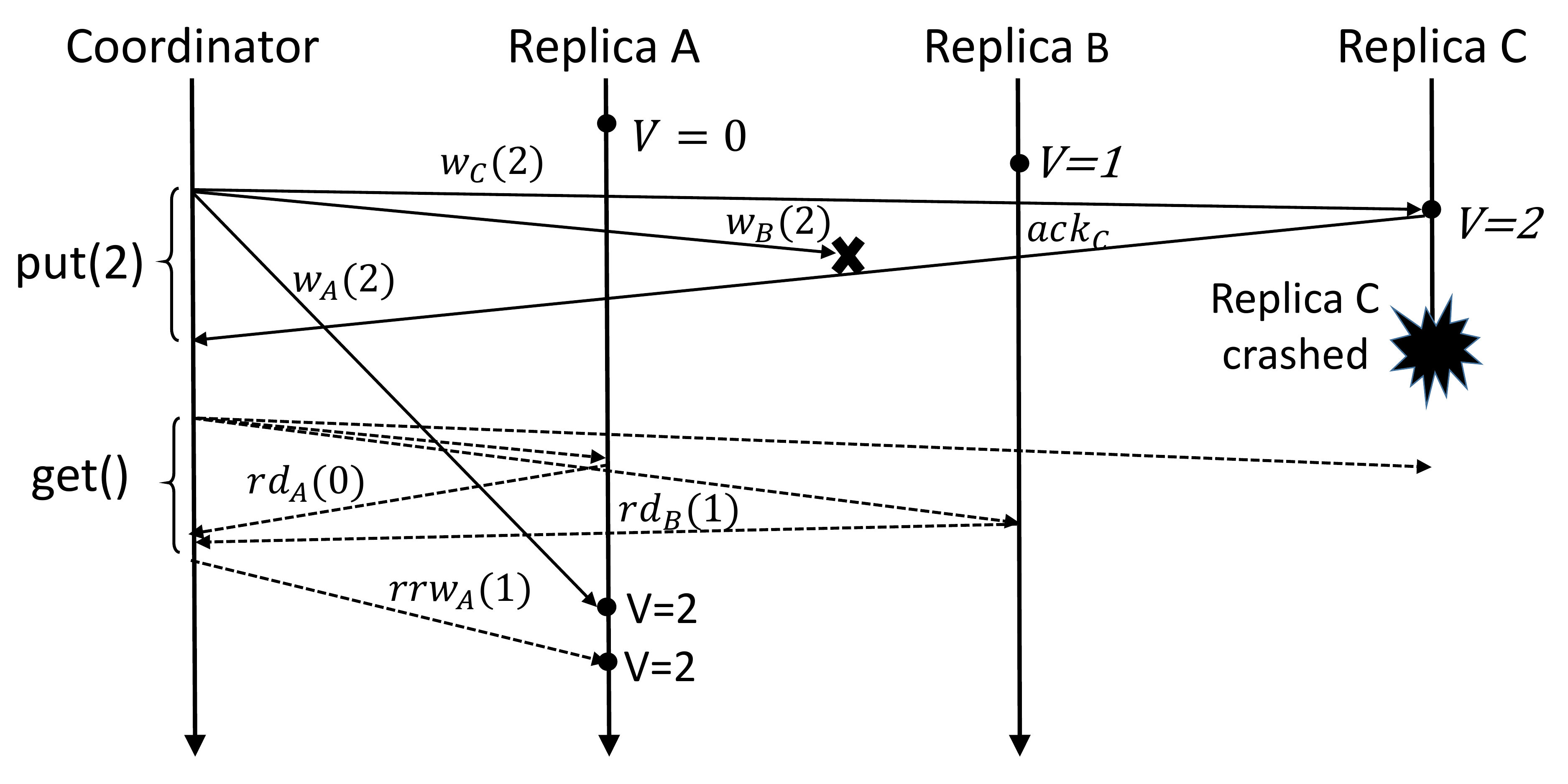}
\caption{A timeline showing that a single read-repair operation does not guarantee convergence of live replicas. In the figure, $w_{r}$ is a write messages to replica $r$, $rd_{r}$ is a message from replica $r$ to the coordinator on the read path, and $rrw_{r}$ is the read-repair message to replica $r$. Time in the figure advances from top to bottom. The messages along the read(-repair) path are shown as dotted lines and along the write path as solid lines.}
\label{fig:msc1}
\end{center}
\vspace{-0.9cm}
\end{figure}

It is commonly believed\footnote{\url{http://wiki.apache.org/cassandra/ReadRepair}} that when a read happens, the read repair protocol will repair the live nodes at the time of the read (assuming they stay alive), bringing them to a common state.
We modeled the read-repair protocol and tried to prove this property, but we failed to come up with appropriate
invariants that would ensure the property. This led us to conclude that the property is not always true.
To see why, consider the timeline in Figure~\ref{fig:msc1}. In this scenario, the client issues a put request with the value $2$, which is routed by the coordinator to all three replicas-- $A, B$, and $C$ (via messages $w_{A}(2), w_{B}(2)$, and $w_{C}(2)$). The replica $C$ successfully updates its local store with this value. Consider the case when the write consistency is one and the put operation succeeds (in spite of the message $w_{B}(2)$ being lost and the message $w_{A}(2)$ being delayed). Now assume that the replica $C$ crashes, and the last write (with value $2$) is in \emph{none} of the alive replicas-- $A$ and $B$. If we consider the case where $B$ has the latest write (with value $1$) among these two live nodes, a subsequent read-repair would write the value $1$ read from $B$ to $A's$ store (via message $rrw_{A}(1)$ in Figure~\ref{fig:msc1}). But before this write reaches $A$, $A$ could get a pending message from the network ($w_{A}(2)$) and update its value to a more recent value-- $2$. In this situation, after replica $A$ has updated its value to $2$, the two alive replicas ($A$ and $B$) do not have consistent values. Due to the lack of hints or processes with hints having crashed, $B$ may never receive the later write (message $w_{B}(2)$).
\vspace{-0.5cm}

Based on these insights, we prove a more involved property of read-repair:
\begin{theorem}
After any sequence of reads and writes, if all operations stop\XSpace{\footnote{The assumption
that updates stop coming is part of the original definition of eventual consistency~\cite{bayou}. There are other formalizations without this assumption~\cite{Kaki15}; however, the read-repair protocol does \emph{not} satisfy eventual consistency without it.}}
except for an infinite sequence of reads of a key, then assuming the set $R$ of replicas are alive at the time of the first such read
and thereafter, the replicas in $R$ will eventually converge to the same value.
\label{result2}
\end{theorem}

We prove the above result also using program verification on the program model. Intuitively, as long as an indefinite number of reads to the key happen, the system will ensure that the subset of live replicas
responsible for the key converge to the same value, eventually. A read-repair may not bring the live replicas to sync
if there are some pending messages in the system. However, since there is only a finite amount of \emph{lag} in the system
(pending messages and hints, etc.), and once the system is given enough time to finish its pending work, a read-repair will
succeed in synchronizing these replicas.

\XSpace{It is tempting to think that one could implement any CRDT and reach eventual consistency of the CRDT store using solely read-repair,
similar to the Corollary we obtained for Theorem~\ref{result1}. However, this is tricky when clients send operations to do on the
CRDT and the conflict-resolution in read-repair happens using state-based merges.
For instance, assume that we implement a counter CRDT, where state-merges take the maximum of the counters, and operations increment the counter~\cite{Shapiro11}. Then we could
have the following scenario: there are $7$ increments given by 
clients, and the counter at replica $A$ has the value $7$ and replica B has $5$ (with two increments yet to
reach $B$), and where a read-repair merges the values at these replicas to $7$, after which the two pending increments arrive at $B$ incrementing it to $9$ (followed by another read-repair where B also gets updated to $9$).
Note that consistency is achieved (respecting our Theorem~\ref{result2}), but the counter stores the wrong value.

Systems such as Riak implement CRDTs~\cite{riak-crdts}
using these underlying protocols by \emph{not} propagating operations (like increments) across replicas, but rather increment one replica, and pass the \emph{state} to other replicas, and hence implement a purely state-based CRDT~\cite{RiakCRDT}.}

\section{Program Models for the Protocols}
\label{sec:model}

\begin{figure}[ht]
\begin{center}
\includegraphics[width=16cm, height=6cm]{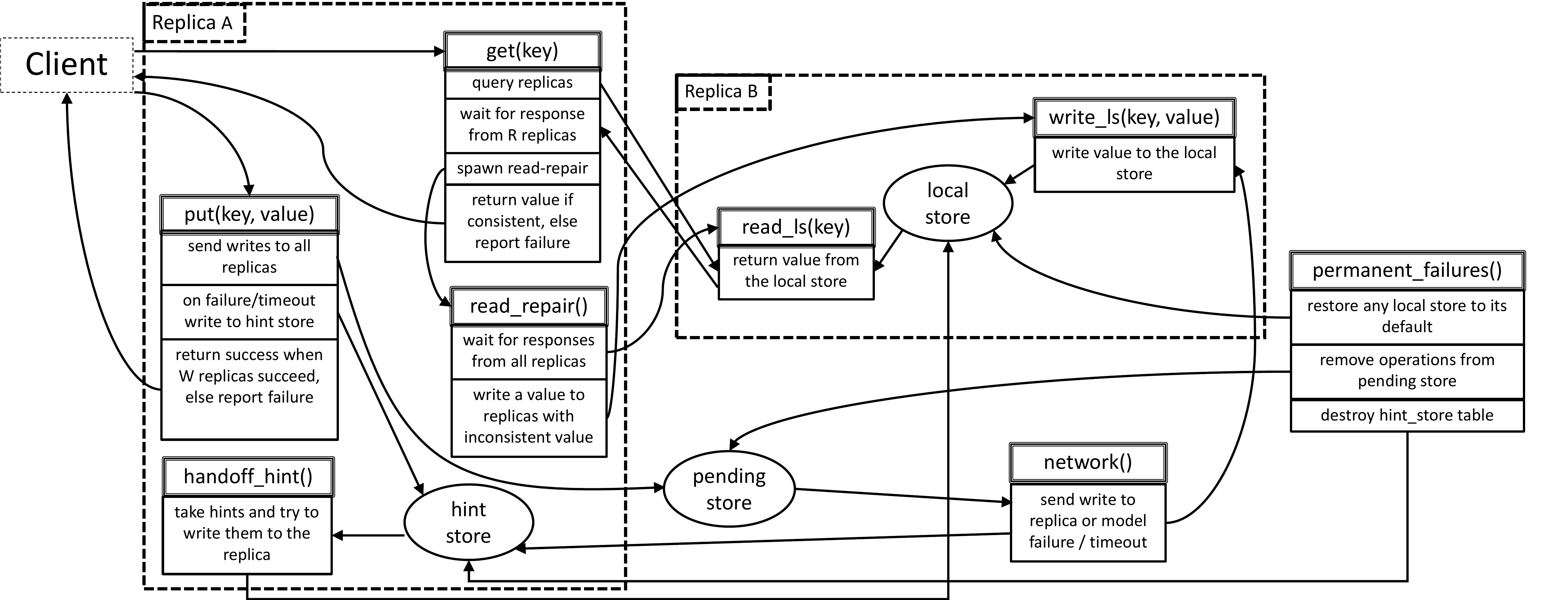}
\caption{Architecture of the model with boxes for functions, ellipses for data structures and arrows for communication.}
\label{fig:arch}
\end{center}
\vspace{-1cm}
\end{figure}

In this section we describe how we model the anti-entropy protocols used in eventually consistent key-value stores.
The architecture of our model is depicted in Figure~\ref{fig:arch}.

\subsection{Program Model Overview}
Our C program model represents replicas as concurrently executing threads that communicate by asynchronous message passing. Each replica uses several functions (\textit{get}, \textit{put}, \textit{write\_ls}, \textit{read\_ls}, etc.) to update their state and communicate. Replica state is kept in globally shared array data structures (\textit{local\_store}, \textit{hint\_store}, etc.). Furthermore, in order to model asynchronous messaging, we maintain a data structure \textit{pending\_store}, that represents messages in the network that have not yet been delivered. The functions in our model include:
\begin{itemize}
 \item The \textit{get} and \textit{put} functions at coordinators that forms
  the interface to clients for reading and writing key-values.
 \item An internal function \textit{handoff\_hint} for each replica that runs all the time and removes hints from the hinted-handoff table and propagates them to the appropriate replicas (provided they are alive).
 \item An internal function \textit{read\_repair} which is part of the read path, waits for all the replicas to reply, and on detecting replicas with inconsistent values writes the consistent value to those replicas.
 \item Internal functions \textit{read\_ls} and \textit{write\_ls}, that read from and write to the local stores (provided they are alive).
 \item An internal function \textit{network} that is invoked repeatedly and delivers messages in the pending store to replicas.
 \item An internal function \textit{permanent\_failures}, which when permanent failure is modeled is invoked repeatedly, and can remove elements from the pending set (modeling loss of messages), restore any local store to its default value (modeling store crashes), and destroy hinted-handoff tables.
\end{itemize}

Note that modeling these function using fine-grained concurrency ensures the possibility of
arbitrary interleaving of function invocations as well as arbitrary delays.
Also, transient failures, where nodes fail but resume later with the correct
state, can be seen as delays in processes, and hence are captured in this
concurrency model. The thread that delivers messages in the pending set captures arbitrary delays in the network.
The \textit{read\_ls} and \textit{write\_ls} functions are modeled abstractly
as idempotent CRDTs by defining them as stubs which maintain specified properties. When
testing, these functions need to be instantiated to particular conflict-resolution strategies (e.g., MV or LWW).

When a client issues a get request for a key in our model, the request is routed to the coordinator that is determined for this key according to an abstract map (hence capturing all possible hashing schemes). Every key-value pair is replicated across multiple nodes, where the number of nodes that contain the key-value pair is determined by a replication factor. The coordinator maintains a preference list of replicas that contain data values for keys that are mapped to it. Along the read path, the coordinator asynchronously issues the read request to all replica threads (an asynchronous call to a replica is depicted in Figure~\ref{fig:arch} as an arrow from the \textit{get} function to \textit{read\_ls}). As shown in Figure~\ref{fig:arch}, the coordinator blocks for a non-deterministic amount of time or until it receives enough responses (the arrow directed from \textit{read\_ls} to \textit{get}) as specified by the read consistency level $R$. After receiving responses from $R$ replicas, it returns the read value(s) to the client. If read-repair is enabled, the coordinator also spawns a background thread (depicted as a call to \textit{read\_repair} from \textit{get} in Figure~\ref{fig:arch}) which will wait for responses from the other replicas \XSpace{(it already knows about responses from the $R$ replicas)} for a non-deterministic amount of time. This thread determines the most recent data value of all the values stored in the various replicas, and writes it to the replicas with stale values.

When a client issues a \textit{put} request to store a key-value pair, the request is routed to the appropriate coordinator. The coordinator asynchronously issues write requests to all replica threads in its preference list. The coordinator then blocks for a non-deterministic  amount of time or until it receives enough responses, as specified by the write consistency level $W$. To model arbitrary network delays and replica failures, write operations to these replicas are inserted by the coordinator into the pending store (in Figure~\ref{fig:arch}, this is depicted as an arrow from \textit{put} to \textit{pending\_store}).
If the coordinator receives responses from $W$ replicas, it informs the client about the successful \textit{put} operation.

A background \textit{network} thread models arbitrary network delays and failure scenarios as it removes writes operation from the pending store data structure and either updates the local store of the appropriate replica with the write or simply loses the operation.
When the hinted-handoff protocol is enabled and read-repair is disabled, we assume that the write operations are not lost. In this scenario, when losing/removing the write operation from the pending store, the \textit{network} thread inserts the operation as a hint in the hinted-handoff table of the appropriate coordinator.
The \textit{permanent\_failures} thread does not execute in this case and data in the global data structures is not lost.

\subsection{Program Model Testing and Validation}
\label{sec:test}

We tested our program model to make sure that it corresponds to actual systems. For our tests, we implemented the stubs for model failure and non-determinism in message arrivals. In particular, we used random coin-tosses instead of non-deterministic choices as in the verification model. We also provided concrete implementations for conflict-resolution strategies for operations on CRDTs in the form of LWW and MV. We then wrote a test harness that arbitrarily issues put and get operations for various key-value pairs. We then checked that the results of these operations could be realized by the actual eventually consistent key-value stores. We also used \chess~\cite{Musuvathi08}, which is a systematic testing tool for concurrent programs, to systematically enumerate all possible thread schedules. Using \chess, we were able to ensure that our model realized strange but possible behaviors of the eventually-consistent stores.

We exhaustively tested many scenarios. Here, we discuss a configuration with three replicas, where the write consistency level is set to two, and the read consistency level is set to one. One interesting scenario is  where the client successfully performs a write operation on a key with a value $0$, followed by an unsuccessful write on the same key with a value $1$. A subsequent read of the key returns the value $1$. This is a counterintuitive scenario, but it can manifest in a real system because failures are not guaranteed to leave the stores unaffected and an unsuccessful write can still write to some of the replicas.
In another scenario, the client successfully performs two consecutive write operations to a key with values $0$ and $1$. Subsequently, one read returns the value $1$, while a subsequent read returns the stale value $0$. This behavior can happen in a real system where the client gets staler values over time. In particular, this scenario occurs when the two replicas store the value $1$ after the second write operation (remember the write consistency level is two) and the third replica still stores the stale value $0$.
Finally, we consider a scenario where there are four consecutive successful writes to a key with values $0$, $1$, $2$, and $3$. (As above, consider a configuration with three replicas but where both read and write consistency levels are set to one.)  If the subsequent two reads for the same key return values $2$ followed by $1$, then a following third read cannot return the value $0$. This scenario cannot happen because the three replicas must have values $1$, $2$, and $3$ at the time of the last read (the reader is invited to work this case out on paper). We used \chess to confirm the realizability of the first three scenarios, and the infeasibility of the last scenario.\XSpace{ \chess took from less than a second to up to 10 minutes to exhaustively explore all interleavings corresponding to these four test harnesses.} We were also able to observe some of these scenarios in a real installation of Cassandra.
\vspace{-0.3cm}

\section{Verification of the Anti-entropy protocols}
\label{sec:verification}
In this section, we describe our verification methodology, and our verification of the hinted-handoff and read-repair anti-entropy protocols using the program model. We use the \emph{deductive verification} style for proving programs correct.
For sequential programs, this style is close to Hoare logic style reasoning~\cite{Apt81}.
It proceeds by the programmer annotating each method with pre/post
conditions and annotating loops with loop invariants with desirable program properties. Furthermore, in order to prove that functions terminate, the user provides \emph{ranking functions} for loops (and recursive calls) that map states to natural numbers and must strictly decrease with each iteration. Reasoning that annotations are
correct is done \emph{mostly} automatically using SMT solvers\XSpace{, with very little help from the user}.

There are several different approaches to verify concurrent programs, especially for modular verification.
We use the \vcc tool~\cite{Cohen09} to verify our models. \vcc is a verifier for concurrent C programs\XSpace{~\footnote{Even though our model and invariants apply to unbounded number of instances, verification of C programs, strictly speaking, assumes integer manipulations to \texttt{MAX\_INT} (i.e., typically $2^{32}$ on 32-bit architectures).}}. The basic approach we take to verify our models
is to treat each concurrent thread as a sequential thread for verification purposes, but where every access to a shared variable is preceded and succeeded by a \emph{havoc} that entirely destroys the structures shared with other threads. However, this havoc-ing is guarded by an \emph{invariant} for the global structures that the user provides. Furthermore, we check that whenever
a thread changes a global structure, it maintains this global invariant.
This approach to verification is similar to \emph{rely-guarantee} reasoning~\cite{Jones83},
where all threads rely and guarantee to maintain the global invariant on
the shared structures.

Another key aspect of the verification process is writing the \emph{specification}.
Though the specification is written mainly as assertions and demanding that certain functions terminate, specifications are often described accurately\XSpace{ and naturally} using \emph{ghost code}~\cite{Apt81,Cohen09}. Ghost code is code written purely for verification purposes (it does not get executed) and is written as instructions
that manipulate ghost variables. It is syntactically constrained so that real code can never see the ghost state. Hence this ensures that the ghost code cannot affect the real code. We use ghost code to model the taint-based specification
for eventual delivery (see Section~\ref{subsec:hh}). It is important that the protocol does not see the tainted write, because we do not want a flow of information between the executable program and the specification. We also use ghost code to maintain abstractions of concrete data structures (e.g., a set abstracts an array).

We performed the program model verification on an Intel Core i7 laptop with 8 GB of RAM, running Windows 8 and using Visual Studio 2012 with \vcc v2.3 as a plugin. Our model\footnote{The code is available at \url{https://github.com/palmskog/evcons-model}} consists of about 1500 lines of code and annotations, where about 900 lines are executable C code and the rest are annotations not seen by the C compiler. \XSpace{The annotations comprise ghost code (20\%) and invariants (80\%). The total time taken for the verification of the whole model is around a minute.}

\XSpace{We extensively used testing, especially in early stages, to assert invariants
that we believe held in the system at various points in the code. Prior to verification, which requires strong inductive invariants, testing allowed
us to gain confidence in the proof we were building (as well as the model
we were constructing). These invariants then were the foundation on which
the final proof was built upon.}

\vspace{-0.3cm}

\subsection{Verifying the Hinted-handoff Protocol}
\label{subsec:hh}
As explained in Section~\ref{sec:main-result}, verification that the hinted-handoff protocol maintains strong eventual consistency under transient failures and for idempotent operation-based CRDT reduces to verification of eventual delivery. Recall that eventual delivery is the property that every successful write eventually gets delivered to every replica at least once.

We model eventual delivery using a ghost field \emph{taint}, that records a particular (exactly one) write operation issued to the coordinator. We assert that this taint will eventually propagate to each replica's local store. Intuitively, the write that was chosen to be
tainted will taint the value written, and this taint will persist as the
value moves across the network, including when it is stored in the hint store
and the pending store, before being written to the local store.
Taints persist and do not disappear when they reach the local store.
Hence, demanding that the local stores eventually get tainted captures the
property that the chosen write is eventually delivered at least once to every local store.
Note that the tainted values are system-agnostic ghost fields, and hence proving the above property for an arbitrary write ensures that \emph{all writes} are eventually delivered.

To prove the specification, we introduce 3 ghost fields: (a) \textit{ls\_tainted\_nodes}, the set of replicas that have updated their local store with the tainted write, (b) \textit{hs\_tainted\_nodes}, the set of replicas for which the coordinator has stored the tainted write operation as a hint in its hint store, and (c) \textit{ps\_tainted\_nodes}, the set of replicas for which the tainted write has been issued, but where its delivery is pending in the network.

We add ghost code to maintain the semantics of the taint in various functions,
including \textit{put}, \textit{network} and \textit{handoff\_hint}.
Every time any of these functions transfers values, we ensure that the taints
also get propagated. When a value is written to a local store, the store is considered tainted if it either already had a tainted value
or the new value being written is tainted; otherwise, it is untainted.
\XSpace{(In fact, the taint-based store can itself be seen as an operation based CRDT
 which never loses the taints.) }Furthermore, we add ghost code to update the ghost fields described above.

For eventual delivery, we want to prove that, when all replicas remain available and all the read/write operations have stopped, the tainted write operation is eventually propagated to the local stores of all the replicas.
We prove eventual taintedness of stores by proving a global \emph{termination property}.
We model the point where inputs stop arriving using a variable $\textit{stop}$ and by making all nodes alive while
disabling the functions $\textit{get}$ and $\textit{put}$. We then prove, using a
ranking function, that the model will eventually reach a state where all nodes corresponding to
the tainted key are tainted.
We first specify a (safety) invariant for the shared state and specify a ranking function
on the state for arguing eventual taintedness. The invariant of the shared state asserts that for every replica
responsible for the tainted key, either its local store is tainted or there is a
tainted write pending in the network for it, or there is a hint in the
corresponding coordinator which has a tainted write for it.
More precisely, for each replica responsible for the tainted key, we demand
that the replica is in one of the ghost sets\XSpace{, namely,
\textit{ps\_tainted\_nodes},  \textit{hs\_tainted\_nodes},
and \textit{ls\_tainted\_nodes}}:

$\forall\, r.~ (~r \in \textit{ps\_tainted\_nodes~} ~\vee ~ r \in \textit{hs\_tainted\_nodes~} ~ \vee ~ r \in \textit{ls\_tainted\_nodes}~),$

\noindent where the quantification is over replicas $r$ responsible for the tainted key.
The ranking function is a function that quantifies, approximately, the \emph{time} it would take for the system to reach a consistent state. In our case, the ranking function $| \textit{hint\_store} | +$ $2\cdot|\textit{pending\_store}|$ suffices. We prove
that the rank decreases every time the function executes.
\XSpace{Intuitively, a taint that is pending in the network may take two steps to get to the local store, since it may first be transferred
to the hint store and then to the local store, while tainted messages in the hint store take one step to the local store.}
Finally, we assert and prove that when the rank is zero, all nodes corresponding to the tainted key are tainted.

\vspace{-0.3cm}
\subsection{Verifying the Read-repair Protocol}
As explained in Section~\ref{sec:main-result}, we want to verify that the read-repair protocol maintains eventual consistency in the presence of permanent failures (as stated in Result~\ref{result2} in Section~\ref{sec:readrepair}).
We prove this result both when hinted-handoff is turned on as well as when
it is disabled (we capture whether hinted-handoff is enabled/disabled using
a macro directive, and prove both versions correct).
For simplicity of presentation we only explain here the case when the hinted handoff protocol is disabled.
Recall that permanent failures can (a) modify the local store by setting them to default values, (b) remove an operation from the pending store, and (c) destroy the hint store.

For eventual consistency, we want to prove that when all the write operations have successfully returned to the client, then after only a finite number of read operations on a key, the read-repair mechanism ensures that the set of $R$ available replicas will converge.  When the writes stop and only the read of a particular key occurs\XSpace{ (infinitely often)},
we disallow the scheduling of $\textit{put}$ and disallow $\textit{get}$ on all but the tainted key,
and also bring all nodes alive and disable the methods that model failure of nodes.
We verify eventual consistency by specifying safety invariants and a ranking function.
The main component of the ranking function is the size of the pending store,
$|\textit{pending\_store}|$, which decreases whenever the network executes.
\XSpace{Intuitively, once the pending messages in the network get delivered, the subsequent
execution of $\textit{get}$ will issue a read-repair that will bring the nodes of the
tainted key to convergence.}
\vspace{-0.3cm}

\section{Related Work}
\label{sec:related}

Amazon's use of formal techniques~\cite{Newcombe15} for increasing confidence in the correctness of their production distributed systems is in the same spirit as our work.\XSpace{ The engineers at Amazon have successfully used TLA+~\cite{tla} to formally specify the design of various components of distributed systems.}
Like our programs, TLA+ models by Amazon engineers are executable and can be explored using model checking to uncover subtle bugs.
\XSpace{The formal TLA+ specifications are executable like our program models and these design specifications have been systematically explored using a model checker to uncover several subtle bugs in these systems.}
\XSpace{However, instead of modeling distributed algorithms in TLA+, we model them as C programs. }Newcombe et al.~\cite{Newcombe15} acknowledge that modeling systems in a high-level language like C, as we do, increases the productivity of the engineers. More importantly, in addition to model checking up to certain trace lengths, a model written in C lends itself to mostly automated verification using tools like \vcc that utilize automated constraint solvers, and that can verify unbounded instances of the system.
There have also been efforts towards formally modeling key-value stores like Cassandra using rewriting logic in Maude~\cite{maudeCass}. However, this model checking approach is either not exhaustive or is exhaustive on bounded instances, while ours is exhaustive on unbounded instances.
Recently, there has been work on programming languages that ease development of distributed systems, in particular, with respect to consistency properties at the application level~\cite{Burckhardt-ECOOP-12} and
fail-free idempotence~\cite{Ramalingam13}.  Also, work by Sivaramakrishnan et al.~\cite{Kaki15} introduces a declarative programming model for eventually consistent data stores, that includes a \emph{contract} language that can express fine-grained application level consistency properties.\XSpace{ Lesani et al.~\cite{Lesani16} developed a framework for for modular verification of causally
consistency for replicated key-value stores.}
Hawblitzel et al.~\cite{Hawblitzel15} use TLA-style refinement to verify an implementation of Paxos consensus in Dafny, while Woos et al.~\cite{Woos2016}, verify an implementation of the Raft consensus algorithm in the Coq proof assistant; both approaches capture node and network behavior in a programming language setting, like we do, but focus on top-down system development.
Burckhardt et al~\cite{Burckhardt-POPL-14} explore logical
mechanisms for specifying and verifying properties over replicated data types,\XSpace{ In particular, Burckhardt in his book~\cite{Burckhardt-Book} gives an extensive treatment of the principles of consistency of distributed systems.} and propose a framework for specifying replicated data types using relations over events and verifying their implementation using replication-aware simulations. 
\XSpace{Deductive verification using automatic tools, such as VCC~\cite{Cohen09} have been extensively used for verifying systems in domains other than distributed systems.}\XSpace{Some of the examples are: verifying a hypervisor for the isolation property~\cite{vcc-hypervisor}, verifying operating systems like Verve~\cite{verve} and ExpressOS~\cite{Mai13} for security, verifying the L4 microkernel for functional correctness~\cite{sel4} and verifying high-level applications for end-to-end security~\cite{Ironclad}. Recently, a blend of TLA-style state-machine refinement and Hoare-logic verification was used in \cite{Hawblitzel15} for verification of distributed systems.}
\vspace{-0.3cm}

\section{Conclusions}
\label{sec:conclusion}
We have shown how formal properties of production key-value (NoSQL) storage systems can be inferred by finding properties
that are provable for a high-level distributed protocol model modeling the implementation.
Furthermore, we have proposed a modeling technique using programs in concurrent C, which gives executability of models, testing, and
mostly automated formal verification using the \vcc tool.
We have verified both eventual delivery of the hinted-handoff protocol under transient failures as well as eventual consistency of the read-repair protocol when arbitrary
number of reads are issued. We also discovered several surprising counterexamples during the verification for related conjectures based on online documentation, and this experience helped us develop a firm understanding of when and how these protocols guarantee eventual consistency. To the best of our knowledge, this is the first time these anti-entropy protocols have been verified exhaustively using deductive verification.
%
We believe the methodology proposed in this work is promising and applying our methodology to a larger class of production protocols (e.g., blockchain, Google's Cloud Spanner) is interesting future work.


\bibliographystyle{splncs}
\bibliography{refs}

\end{document}